\documentstyle[graphicx]{mn}

\def\mpc{$\,h^{-1}\,$Mpc}
\def\%{~per~cent}
\def\lsim{\lower.73ex\hbox{$\sim$}\llap{\raise.4ex\hbox{$<$}}$\,$}
\def\gsim{\lower.73ex\hbox{$\sim$}\llap{\raise.4ex\hbox{$>$}}$\,$}

\title{The Local Hole in the Galaxy Distribution: New Optical Evidence}
\author[Busswell et al.]
{
G.S. Busswell, T. Shanks, P.J. Outram, W.J. Frith, N. Metcalfe \& R. Fong  \\
Department of Physics, University of Durham, Science Labs, South Road, Durham DH1 3LE, United Kingdom}

\begin{document}

\maketitle

\begin{abstract}
We present a new CCD survey of bright galaxies predominantly within the Northern
and Southern strips of the 2dF Galaxy Redshift Survey (2dFGRS) areas. We use the new CCD
data to check the photographic photometry scales of the 2dFGRS, APM Bright Galaxy Catalogue, 
APM-Stromlo Redshift Survey, Durham-UKST (DUKST) survey, Millenium Galaxy Catalogue (MGC) and Sloan Digital 
Sky Survey (SDSS). We find evidence for scale and zero-point errors in the 2dFGRS northern field, 
DUKST and APM data of 0.10, 0.24 and 0.31 mag. respectively; we find excellent agreement 
with the MGC and SDSS photometry. We
use our CTIO data to correct the photographic photometry, we then compare the CCD
number counts in both the Northern and Southern survey areas. We find conclusive
evidence that the Southern counts with B$<$17 mag are down by $\approx$30\%
relative to both the Northern counts and to the models of Metcalfe et al in the
same magnitude range. We further compare the number redshift distributions from
the B$<17$ mag Durham-UKST and B$<$19.5 2dFGRS redshift surveys using the
corrected photometry. While the Northern n(z) from 2dFGRS appears relatively
homogeneous over its whole range, the Southern n(z) shows a 30\% deficiency out
to z=0.1; at higher redshifts it agrees much better with the Northern n(z) and
the homogeneous model n(z). The Durham-UKST n(z) shows that the Southern `hole'
extends over a  20$\times$75 deg$^2$ area. The troughs with z$<$0.1 in the Durham-UKST n(z)
appear deeper than for the fainter 2dFGRS data. This effect appears to be real since
the troughs also appear to deepen in the 2dFGRS data when magnitude limited at B$<$17 mag and
so this may be evidence that the local galaxy distribution is biased on \gsim50\mpc\  scales 
which is unexpected in a $\Lambda$CDM cosmology. Finally,
since the Southern local void persists over the full area of the APM and
APM Bright Galaxy Catalogue with a $\approx$25\% deficiency in the counts below B$\approx$17, 
this means that its extent is approximately 
300~$h^{-1}$Mpc$\times$300~$h^{-1}$Mpc on the sky as well as $\approx$300~$h^{-1}$Mpc in the redshift direction.
Such a 25\% deficiency extending over $\approx$10$^7$~$h^{-3}$~Mpc$^3$ may imply that the
galaxy correlation function's power-law behaviour extends to $\approx$150~$h^{-1}$Mpc
with no break and show more excess large-scale power than detected in the 2dFGRS correlation
function or expected in the $\Lambda$CDM cosmology.   

\end{abstract}


\section{Introduction}

It has been apparent for many years that the B galaxy number counts may be
steeper at bright magnitudes than expected in a homogeneous, unevolving,
cosmological model (Shanks et al. 1990). The counts appeared to have a
Euclidean slope  whereas the models predict a much flatter slope due to the  
expansion and the K-correction. Shanks et al. suggested
that these steep counts maybe caused by large-scale galaxy clustering
whereas other authors such as Maddox et al. (1990a) tended to favour the suggestion that
the steepness was due to evolution. Both ideas had difficulties - 
the unevolved n(z) at B=17 and B=21 argued against galaxy luminosity evolution, while 
the local void would have to persist to almost 300$h^{-1}$Mpc if the low number counts were caused by
galaxy clustering.

Most of these authors used counts made from measurements of photographic plates.
This meant that there remained the possibility that these observations were in
error, due to non-linearities in the photographic magnitude scales. More recently
it has become possible to survey large areas of the sky with linear CCD
detectors. At first these were used to calibrate the photographic magnitudes but
this effort has been hindered by the quite large rms errors in the photographic
photometry found even at relatively bright magnitudes. The CCD detectors are now
large enough to provide the survey data by themselves and in this paper we   
describe a new survey primarily in the 2dF galaxy redshift survey areas, using
the CTIO Curtis Schmidt CCD camera.

The advent of deeper galaxy redshift surveys such as 2dFGRS opens up a new
possibility of investigating the scale at which the local hole ceases to  
dominate and the  Universe appears to become homogeneous.
In previous surveys such as the Durham-UKST Galaxy
Redshift survey, the n(z) extended to z$\approx$0.1 and it was found that count
models which quite successfully fitted galaxy counts at B$>$18 did not fit the
B$<$17 n(z) even in the range 0.05$<$z$<$0.1 which might have been thought to be
safely outside any local inhomogeneities. The fact that the model n(z)
over-predicted the observed n(z) all the way to z=0.1 meant that if the local
void hypothesis was correct then the `hole' in the galaxy distribution had to
extend to z$\approx$0.1 or r$\approx$ 300\mpc.

In this paper we first present details of our observations and data reduction in Section 2. In
section 3 we use our CTIO CCD data to check the calibration of several previous galaxy number 
counts so that the corrected results from the various authors can be compared to the CTIO counts 
themselves. In section 4 the CTIO counts are presented in each band. The number redshift 
distributions from the 2dF Galaxy Redshift Survey (2dFGRS; Colless et al. 2001) and the Durham-UKST 
Redshift Survey (Ratcliffe et al. 1998) are examined in Section 5 to investigate the possible existence
of a `local hole' in the galaxy distribution in the Southern Galactic Cap (SGC). 
A discussion follows in Section 6.

\section{Data Reduction}

\subsection{Observations}

The observations were taken using the 0.6m Curtis Schmidt Telescope at CTIO, La Serena, Chile.
We had two filters, Harris B and R, with magnitude limits of 20.5 and 19.5 respectively,
and the imaging data was taken over
2 observing runs of 7 nights, each in excellent weather conditions. The North Galactic Cap (NGC) observations
were taken from 8-14 April 1999 inclusive (observers GSB and TS), and the data in the South Galactic 
Cap (SGC)
from 17-23 October 2000 inclusive (observers GSB and PJO).  The CCD is a 2048x2048 24 micron chip with 2.3
arcsecond pixels, so that when the
bias was subtracted this resulted in a 1.69 degree$^2$ field of view for each exposure.

In the NGC, we observed 3 main strips of sky in B and R at declinations of
0, -5 and -10 degrees, where the equatorial and -5 degree strips overlap with the 2dF Galaxy Redshift
Survey (Colless et al. 2001).  The strips were
1.3 degrees wide, which was dictated by the field of view of the telescope, and had an RA range from
9hrs. 45mins. up to 15hrs.  The strips were joined
at the ends by two smaller strips which were constant at declinations of 10hrs and 15 hrs.
In total, this gave a potential 300 hundred square degrees
of data in our B and R filters, assuming we had photometric conditions throughout all 7 nights.

In the SGC, we again observed in 3 main strips at constant declinations, this time at -28, -30 and -32,
with the knowledge that this would completely overlap with the 2dFGRS fields.
The ends of the strips were at RA's of 21hrs. 40mins. and 03hrs. 15mins.  Because the strips
were much closer together than in the NGC, it was possible
to connect the strips by simply doing single exposures every half hour of RA at declinations of -29 and
-31 degrees.  In addition to these three long strips of $\sim$ 100 square degrees each, we also
observed a shorter strip  at a declination of -45 degrees
from an RA of 02hrs. to 02hrs. 45mins. giving an area of $\sim15$ square degrees.  In total this gave
us an imaging area in the SGC of 337 square degrees.

The format of our observations would be to take two standard star frames of suitable Landolt equatorial
fields in each band at the beginning,
the middle and the end of each night.  We would then begin the observations in our chosen field by taking
an exposure in the R band for 120 seconds and
stop the tracking of the telescope for 1 min 18 seconds (when observing in the North, 1 min. 30 secs. in
the South), so that the sky moved over by
one quarter of the CCD chip. This 1 min 18 seconds was sufficient to let the CCD read out which took $\sim45$
seconds.  An exposure of the same
length of time would then be taken in the B filter, and so on.  The result would be a series of frames at a constant
declination in the sky where each frame overlapped its adjacent frames in the same filter by about half a chip.
We would typically be able to cover
about 90 mins. of RA (30 deg$^2$) using this method before the sky was too far over and we would then move the
telescope and begin again.

The bias was subtracted, images trimmed and bad pixels corrected using the IRAF quadproc package.  Typically, 
five or six dawn and twilight
flat field images were taken in each bandpass.  A master flat field image was produced in each filter for
each night by first using the imcombine routine to median together each of the dawn flats and evening flats 
separately.
The resulting frames
were then averaged to produce our B and R master flats for that night.  By dividing the median-ed dawn and twilight
flats we typically found a 1-2\%
gradient from top to bottom of the resulting frame, implying an error of about 0.005-0.01 to our galaxy photometry
due to this effect.  The frames
for a particular night in each filter were then flat-fielded using the IRAF ccdproc package and the appropriate
flat-field master frame.

\subsection{Photometric Analysis}

The fact that adjacent frames of the same filter overlapped by half-a-chip meant that a detailed analysis of the
photometric conditions
throughout each night could be performed.  The results of this photometric analysis are shown in Table
\ref{table:photo}. Given that the weather
was so good during both observing runs we decided to enforce a relatively strict criterion to determine the
nature of the observing conditions. A night was deemed partially photometric if two or more adjacent frames show
significant offsets ($>0.1$ mag) from each other when performing best fits to the magnitude residuals of the 
stars. 
If a particular night was deemed 
to be ``partially photometric'' then none of the frames were used on the strip where the $>0.1$ magnitude offset was
found.

\begin{table}
\centering   
\begin{tabular}{||l|l|l|l||} \hline
        \multicolumn{2}{||l|}{NGC field} &
        \multicolumn{2}{|l||}{SGC field} \\ \hline
         night    & photometric? & night     & photometric?\\ \hline
          1       & yes          &  1        & yes        \\
          2       & partial      &  2        & yes        \\
          3       & partial      &  3        & yes        \\
          4       & yes          &  4        & partial    \\
          5       & partial      &  5        & yes        \\
          6       & yes          &  6        & yes        \\
          7       & yes          &  7        & no         \\ \hline
\end{tabular}
\caption[\small{A summary of the photometric conditions during each night.}]{\small{By using the fact that the
frames in each filter overlap
adjacent frames by half a chip, we were able to analyse the photometric conditions for each night. This table shows
a summary of this analysis from the two observing runs in the North and South Galactic Caps respectively. In total
our survey areas taken
in photometric conditions were 255 and 297 square degrees for the NGC and SGC respectively.}}
\label{table:photo}
\end{table}

In order to ensure accurate zero-point calibration our strategy was to scale each frame based on its
airmass relative to the first frame
in the sequence.  We found that, assuming photometric conditions, the cumulative magnitude offset across a whole
sequence was consistent with the
airmass variation from the first to the last frame.  Since we enforced strict criteria on whether frames had been
observed in photometric conditions, this assumption was justified.  This also meant that the calibration
of each frame was completely independent of all the others in the sequence, eliminating the possibility of the
propagation of magnitude offset errors.
To take these airmass variations into account the frames were scaled according to the equations:

\begin{equation}
f_{B}=10^{0.209(X_{1_{B}}-X_{n_{B}})/2.5}
\end{equation}

\begin{equation}
f_{R}=10^{0.108(X_{1_{R}}-X_{n_{R}})/2.5}
\end{equation}

\noindent where $X_{1_B}$, $X_{n_B}$ are the air-masses in the first and nth frames in a B filter sequence and 
$X_{1_R}$, $X_{n_R}$ the corresponding
air-masses for the R filter.  $f_B$ and $f_R$ are the scale factors for the B and R filters respectively
and the values 0.209 and 0.108 are the quoted airmass coefficients in B and R for the CTIO observatory. By using 
high-mass standard star frames in the
B and R filters we obtained values for the airmass coefficients of 0.19 and 0.10 respectively, in good
agreement with the quoted values.

\subsection{Standard Stars and the Colour Equation}

Once the internal zero-points had been calibrated a global zero-point for each sequence had to be calculated.
We did this by using the B and
R magnitudes of the Landolt (1992) standards stars from our standard star frames which were taken at the
beginning, middle and the end of each night using the fields SA101, SA107, and SA110.  Our standard star exposure
times were 10 secs. in R and 20 secs.
in B.  The IRAF fitparams routine was then used to determine the best fit zero-point offsets (b$_1$ and r$_1$) and
colour-term coefficients (b$_3$ and r$_3$) for each band in the equations:

\begin{equation}
m_b = B+b_1+b_2*X_b+b_3*(B-R)
\label{equation:standard_b}
\end{equation}

\begin{equation}
m_r = R+r_1+r_2*X_r+r_3*(B-R)
\label{equation:standard_r}
\end{equation}

\noindent where B and R are the Landolt standard star magnitudes, X$_b$ and X$_r$ are the air-masses of the
standard star frame for the B and R bands
respectively and m$_b$, m$_r$ are the calculated magnitudes of the stars.  The parameters b$_2$ and r$_2$ are
equal to 0.209 and 0.108 respectively and,
as has been mentioned, are the quoted airmass coefficients for the CTIO observatory.  We found typical values of
0.05 for b$_3$, the colour-term coefficient,
but the error on this value was 0.04.  Since this was comparable to b$_3$, which was itself small, we decided only
to fit b$_1$ in the colour equation.
Because of the same reasons we only fitted the R band zero-point offset, r$_2$ in equation \ref{equation:standard_r}.
Since we did not use any colour
terms in equations \ref{equation:standard_b} and  \ref{equation:standard_r} all reduced  magnitudes of our final
sources will be in B and R$_{kc}$, as used
by Landolt (1992).  The mean of the b$_1$'s and r$_1$'s was calculated over all seven nights and our resulting colour
equations were then:

\begin{equation}
m_b = B+4.533(^+_-0.0053)+0.209*X_b
\label{equation:col_equation_b}
\end{equation}

\begin{equation}
m_r = R+4.203(^+_-0.0067)+0.108*X_r
\label{equation:col_equation_r}
\end{equation}

We used the SEXtractor software package (Bertin et al. 1996) to measure the magnitudes of our
sources and perform star/galaxy separation.
The sources were extracted, sequence by sequence, and frame by frame in each sequence.  The seeing
stellar FWHM was first estimated for a particular frame with an initial call of SExtractor and then the objects
were extracted with a second pass,
making use of the calculated stellar FWHM.   We used the MAG\_BEST parameter in SExtractor for the source
magnitudes and since we had at least two
observations of each source because of our frame overlaps, the mean of the two magnitudes was used in order
to minimize errors in the photometry.
A cosmic ray was  defined as being a source which appeared on one frame but on neither of the adjacent frames
and these would not be included in our
source catalogue, but written to a cosmic ray file.

\subsection{Star/Galaxy Separation}

Star/Galaxy classification was carried out using the SExtractor software. The parameter CLASS\_STAR is assigned a
particular value for a source which varies between 0 and 1 for galaxy and star-like sources respectively. For 
B$<$18 and R$<$17 the SExtraxtor CLASS\_STAR parameter is a good separator of stars and galaxies with 91\% in B
and 90\% in R of sources with either CLASS\_STAR$>$0.9 or $<$0.1. The fact that we have two filters is useful since 
we get four attempts to classify a particular source (as we have overlaps in each filter) instead of two, and 
can consequently achieve lower errors on the CLASS\_STAR parameter. A full account of this technique is given in Busswell (2001).

To verify our star/galaxy separation we have checked our results with other reliable external sources.
We find a galaxy completeness of $>$90\% for B$<$18 when comparing our northern 
equatorial field with $\approx90$deg$^2$ of the currently available Sloan Digital Sky Survey (SDSS; Yasuda et al. 
2001). Stellar contamination is consequently less than 10\% in this magnitude range; the star/galaxy separation 
begins to break down for B$>$18 as the stellar contamination increases. A similar trend is observed in the R-band data 
except the completeness begins to decrease at R=16.5, but is still equal to 84\% for R=17. A similar 
check of the completeness with the Millennium Galaxy Catalogue (MGC; Driver - personal communication) 
in the B-band indicates a 100\% agreement in the source classification for B$<$16, with completeness and
stellar contamination levels of 93\% and 5\% respectively at B=18. It should be noted that the MGC also used 
SEXtractor for the photometry and star/galaxy separation calculations.

\section{CTIO Curtis Schmidt CCD Survey}

\subsection{Photometry Comparisons}

\begin{figure}
\centering
\includegraphics[width=3in,totalheight=3in]{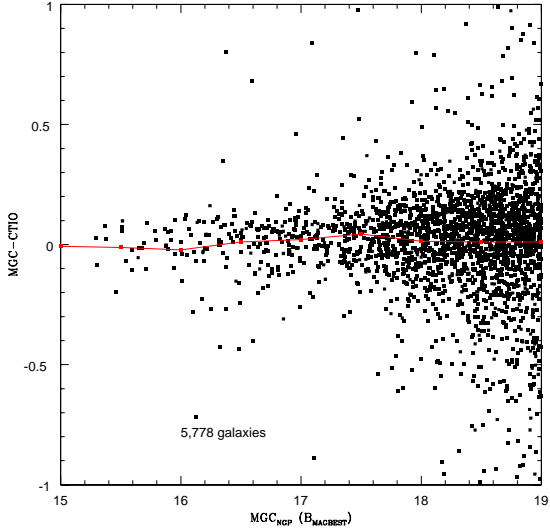}
\caption[\small{Photometry comparison for all common galaxies with the MGC catalogue.}]{Our B-band CCD magnitudes are
plotted against the residual of the MGC magnitudes and our magnitudes for galaxies common to both data-sets. The plot
shows 5,778 galaxies and the black line connects the mean of the residual in each 0.5 magnitude bin.  For B$<$18, we
calculate a mean magnitude difference of MGC-CTIO=0.03 and a $1\sigma$ scatter of 0.1 about this value.  For B$>$18,
our photometry errors  are observed to increase, which is unavoidable given
the length of our exposures and size of our pixels. }
\label{fig:MGC_comp_galaxies}
\end{figure}

In order to check the accuracy of our photometry we have compared our galaxy sample 
with other very accurate data-sets. In Fig. \ref{fig:MGC_comp_galaxies} we show a comparison with the 
B-band MGC data for 5,778
matched galaxies over a 32 deg$^2$ area. The agreement is good, with negligible zero-point or 
scale errors. However, our magnitude errors can be seen to increase significantly for B$>$18. 

\begin{figure}
\centering
\includegraphics[width=3in,totalheight=3in]{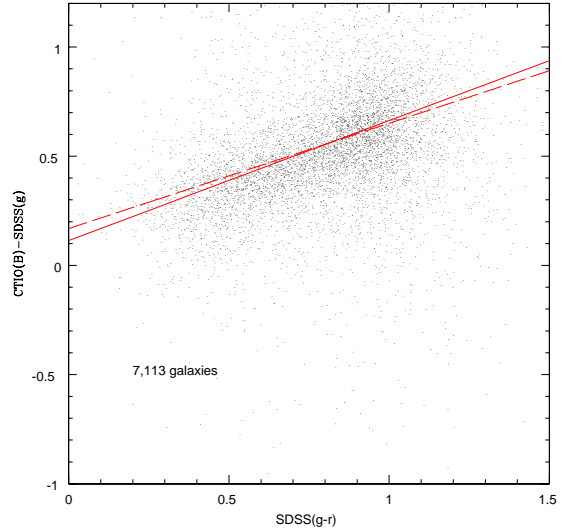}
\caption[\small{Photometry B-band comparison for all common galaxies with the SDSS data.}]{This plot
shows the SDSS ($g^{\star}-r^{\star}$) vs. CTIO(B)-SDSS($g^{\star}$) for 7,133 galaxies in the 90 deg$^2$
overlap region.  The solid line shows the best f
it to the data and corresponds to a colour equation of $B=g^{\star}+0.549(\pm0.07)(g^{\star}-r^{\star})+0.114$
as opposed to the relation $B=g^{\star}+0.482(g^{\star}-r^{\star})+0.169$ (dashed line) quoted in Yasuda et al.
(2001). }
\label{fig:SDSS_comp}
\end{figure}

We can make a similar comparison with the SDSS data (Yasuda et al. 2001) over the 90 square degree overlap region, 
using the g and r filters which need to be converted to our B Landolt band.  The colour equation used in Yasuda et 
al. (2001) is:

\begin{equation}
B=g^{\star}+0.482(g^{\star}-r^{\star})+0.169
\end{equation}

\noindent where $g^{\star}$ and $r^{\star}$ are in the AB magnitude system and the asterisks represent the fact that
the SDSS photometry is preliminary.
By plotting B-g$^{\star}$ against $g^{\star}-r^{\star}$, shown in Fig \ref{fig:SDSS_comp} we can test this colour
transformation for our galaxy sample.  We in fact find the best fit line corresponds to the colour equation 
$B=g+0.549(\pm0.07)*(g^{\star}-r^{\star})+0.114$.
The discrepancy between the colour terms, which is consistent with the $\pm0.07$ error, accounts for the 
0.169-0.114=0.055 difference in the zero-point.  In fact, if we insist on using the 0.482 colour term from 
Yasuda et al. (2001) then the zero-point difference (i.e. the average of the relation   
B-g$^{\star}$-0.482(g$^{\star}$-r$^{\star}$)-0.169 for all common galaxies between the SDSS and our CTIO survey)
is less then 0.01 mag.  The photometry system used for the SDSS data is new and the appropriate transformations 
to other band-passes are relatively poorly understood.  This, combined with the fact that their photometry is still
preliminary, could explain the discrepancy between the colour terms.

\begin{figure}
\centering
\includegraphics[width=3in,totalheight=3in]{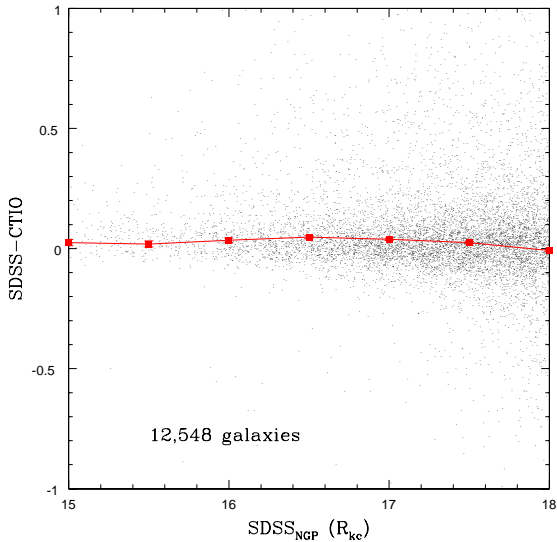}
\caption[\small{Photometry R-band comparison for all common galaxies with the SDSS data.}]{This plot shows a
comparison with the SDSS photometry, but this time in our R Kron-Cousins pass-band.  We have used the colour
transformation, R$_{GKC}$=r$^{\star}$-0.05-0.089(g$^{\star}$-r$^{\star}$), quoted in Blanton et al. (2001)
in order to transform to the R$_{GKC}$ filter.  The relation R$_{GKC}$-R$_{kc}$=0.08 is then used (Shectman
et al. 1996) in order to obtain an SDSS magnitude estimate in our R Kron-Cousins filter.
In this figure we have plotted the SDSS R$_{kc}$ magnitude vs the quantity SDSS(R$_{kc}$)-CTIO(R$_{kc}$)
with the mean of the magnitude difference plotted in half magnitude bins. Excellent agreement is seen over the
magnitude range shown with a mean zero-point difference of SDSS(R$_{kc}$)-CTIO(R$_{kc}$)=0.02($\pm0.01)$ }
\label{fig:SDSS_comp_r}
\end{figure}

A comparison  with the SDSS photometry, this time in the R Kron-Cousins pass-band, is shown in Fig.
\ref{fig:SDSS_comp_r}, where we have plotted the SDSS R$_{kc}$ magnitude estimate against the magnitude difference
SDSS(R$_{kc}$)-CTIO(R$_{kc}$).
The SDSS R$_{kc}$ magnitude is calculated using the colour equation quoted in Blanton et al. (2001) of   
R$_{GKC}$=r$^{\star}$~-~0.05~-~0.089(g$^{\star}$-r$^{\star}$).  A further correction is performed using the
relation R$_{GKC}$-R$_{kc}$=0.08 (Shectman et al. 1996) in order to obtain an estimate of the SDSS R$_{kc}$
magnitude.  We find excellent agreement over the magnitude range shown in Fig. \ref{fig:SDSS_comp_r}, with a   
mean zero-point difference of SDSS(R${kc}$)-CTIO(R$_{kc}$)=0.02($\pm1~0.01)$.

Further checks of our R-band photometry in the NGC and comparisons to our B and R data in the SGC has been carried
out in Busswell (2001) using data from Metcalfe et al. (1998) in the DARS GSA and GNB fields. This data shows an 
intrinsic scatter of 0.06 mags. in both B and R. Mean magnitude differences were found of 
-0.02$\pm0.018$, +0.02$\pm0.02$, and, -0.03$\pm0.02$ for comparisons to R-band data in the NGC and B-band
and R-band data in the SGC respectively.   

\subsection{The APM Bright Galaxy Photometry Correction}

In order to investigate the APM photometry at bright magnitudes we have used our
CTIO SGC data (our un-dust-corrected magnitudes), covering 297 deg$^2$, to check
the accuracy of the publicly available APM Bright Galaxy Catalogue (APMBGC; Loveday et al. 1996)).  
This data reaches a magnitude limit of b$_J$=16.44 and so to reach slightly
fainter magnitudes we have also used the publicly available photometry from the 
APM-Stromlo Redshift Survey (APMSRS), where the magnitude limit is b$_J$=17.15.
We have converted our CTIO data to the b$_J$ magnitude system using both our B 
Landolt and R Kron-Cousins data, in conjunction with the colour equation from   
Pimbblet et al. (2001):

\begin{equation}
b_J=B-0.17(B-R)
\label{equation:kevin}
\end{equation}

\begin{figure}
\centering
\includegraphics[width=3in,totalheight=3in]{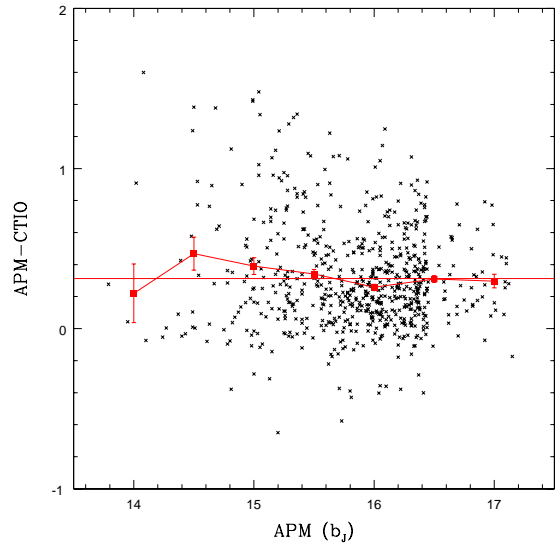}
\caption[\small{Comparison with the APMBGC and APMSRS galaxy photometry.}]{Comparison
with the APMBGC and APMSRS galaxy photometry.  Here we plot APM mag(b$_J$) vs   
APM(b$_J$)-CTIO(b$_J$) for  all the APMBGC and APMSRS galaxies contained in our
297 deg$^2$ area.   The points show the mean of the residual in 0.5 mag. 
bins and the solid line shows the mean zero-point offset over all our measured
magnitudes which we calculate to be 0.313$\pm0.01$.  The low sampling rate of
the APMSRS is  clearly illustrated, as the number of matched galaxies drops
sharply as one moves to fainter magnitudes than b$_J$=16.44, the magnitude
limit of the APMBGC.}
\label{fig:APMcomp}
\end{figure}

Fig. \ref{fig:APMcomp} shows this comparison with the b$_J$ mag. plotted
vs the mag. difference APM-CTIO.  In our 297 deg$^2$ survey area we find 629
matching galaxies with the APMBGC, which are of course brighter than b$_J$=16.44
and 96 matching galaxies with the APMSRS brighter than b$_J$=17.15.  Note that 
the APMSRS has a sampling rate of 1/20 over the full 4300 deg$^2$ APM area and so
we therefore expect to find 20 times more matched galaxies with the APMBGC than
for the APMSRS in a given mag. interval.  This low sampling rate is  clearly   
illustrated in Fig. \ref{fig:APMcomp}, where the number of matched galaxies drops
sharply as one moves to fainter magnitudes than b$_J$=16.44, the magnitude limit
of the APMBGC.  We find a mean zero-point difference of APM-CTIO=0.313$\pm0.01$
mag.   The difference is quite large and we will see in the discussion section   
what this 0.313 mag. correction could mean in terms of the implications for a 
hole in the SGC distribution of galaxies.

\subsection{The Durham/UKST Photometry Correction}

In this section we use our CTIO CCD photometry to check the accuracy of
the Durham/UKST photometry.  The DUKST redshift survey covers 1500 deg$^2$ in 4
strips and our CTIO SGC data overlaps with one of these 4 strips.  The angular
size of the survey is 20$^{\circ}$x75$^{\circ}$ in the SGC to a limiting apparent
magnitude of b$_J$=16.86.  The catalogue consists of $\sim$2500 galaxy redshifts
and therefore by comparing the model of Metcalfe et al. with the DUKST n(z)    
distribution, and taking into account any zero-point discrepancies with our CCD  
data, we can probe a  larger area (1500 deg$^2$) in the SGC than the 2dFGRS can 
($\sim$300 deg$^2$ at present).  Although the DUKST will probe lower redshifts,
the larger area will mean tighter constraints on the size and angular extent of  
the SGC deficiency in the galaxy distribution.

The DUKST galaxies were selected from the Edinburgh/Durham Southern
Galaxy Catalogue (EDSGC: Collins, Heydon-Dumbleton \& MacGillivray 1988).  
Ratcliffe et al. (1998) applied small corrections to
the EDSGC b$_J$ magnitudes in each of the 60 fields in an attempt to put them on
the same zero-point scale of the APM catalogue.  These corrections were
calculated from results of Dalton et al. (1995) with the mean correction over all
60 fields, $<b_J^{APM}-b_J^{EDSGC}>$=-~0.05$\pm0.11$ mag.   It should also be
pointed out that the zero-point corrections were performed at b$_J\sim$19.5 as
the APM is expected to give reliable photometry here.  If this mean offset is to 
be believed in the magnitude range of the DUKST (b$_J<17$), one is assuming there
is no scale error in either the APM or EDSGC photometry.

\begin{figure}
\centering
\includegraphics[width=3in,totalheight=3in]{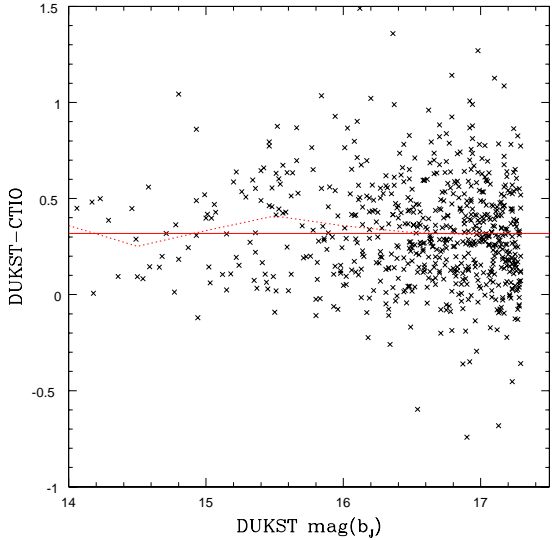}
\caption[\small{Comparison with the EDSGC photometry for the combined set of 28 fields. }]{Here
we plot EDGSC mag(b$_J$) vs EDGSC(b$_J$)-CTIO(b$_J$).  
The dotted line shows the mean of the residual in 0.5 mag. bins and the solid line shows
the mean zero-point offset over all magnitudes which we calculate to be 0.310.  In fact
the mean offset between the EDSGC photometry and the APM data at b$_J$=19.5 is 0.07$\pm0.11$
for these 28 fields which overlap with our CTIO data as opposed to 0.05 for all 60 fields.
This means that we find a 0.31-0.07=0.24$\pm0.012$ zero-point offset between our
CCD photometry and the corrected EDSGC photometry used for the DUKST survey.}
\label{fig:DUKSTallfields} 
\end{figure}

Our CTIO SGC fields overlap with 28 of the 60 EDSGC fields and therefore
we can make extensive tests of a large fraction of the DUKST photometry using our
accurate CCD photometry. To convert to the DUKST b$_J$ magnitude system we use
our B Landolt and R Kron-Cousins photometry in conjunction with equation
\ref{equation:kevin} (Pimbblet et al. 2001). In Fig.
\ref{fig:DUKSTallfields} we plot all the galaxies from each of the 28 fields in
order to determine a mean zero-point offset. The dotted line shows the mean
zero-point offset binned in half magnitude intervals and the solid line shows the
mean zero-point offset over all magnitudes which we calculate to be 0.310.  In
fact the mean offset between the EDSGC photometry and the APM data at b$_J$=19.5 
is 0.07 for the 28 fields which overlap with our CTIO data as opposed to 0.05 for
all 60 fields.  This means that we find a 0.31-0.07=0.24 zero-point offset
between our CCD photometry and the corrected EDSGC photometry used for the DUKST
survey. Although the DUKST photometry was
corrected in order to be consistent with the zero-point of the APM, it is clear
that there are discrepancies compared to our accurate CCD data.  These zero-point
differences appear field dependent although in some fields there are very few
galaxies.

This mean zero-point offset between our CTIO CCD data and that of the   
DUKST may seem surprising given that the DUKST photometry has already been
corrected/brightened by an average value of 0.05 mag. in order that it is
consistent with the APM data at b$_J$=19.5  However, as we have already
mentioned, any scale errors inherent in the APM or the DUKST could mean that this
0.05 correction may not apply at b$_J<$17. Our checks of 
the 2dFGRS photometry (based on the original APM data) showed up no significant  
scale error in the magnitude range $16<$b$_J<19.5$ and therefore we claim that
the problem lies with the EDSGC photometry used for the DUKST survey.  A scale
error of order 0.1 mag/mag in the range 17$<b_J<$19.5 would explain the 0.24 mag.
discrepancy with our CCD data which we have found to be accurate to a few   
hundredths of a magnitude at B$<$17 from checks performed in Section 3.1.

\subsection{Photometry Checks of the 2dFGRS}
 
\begin{figure}
\centering
\includegraphics[width=3in,totalheight=3in]{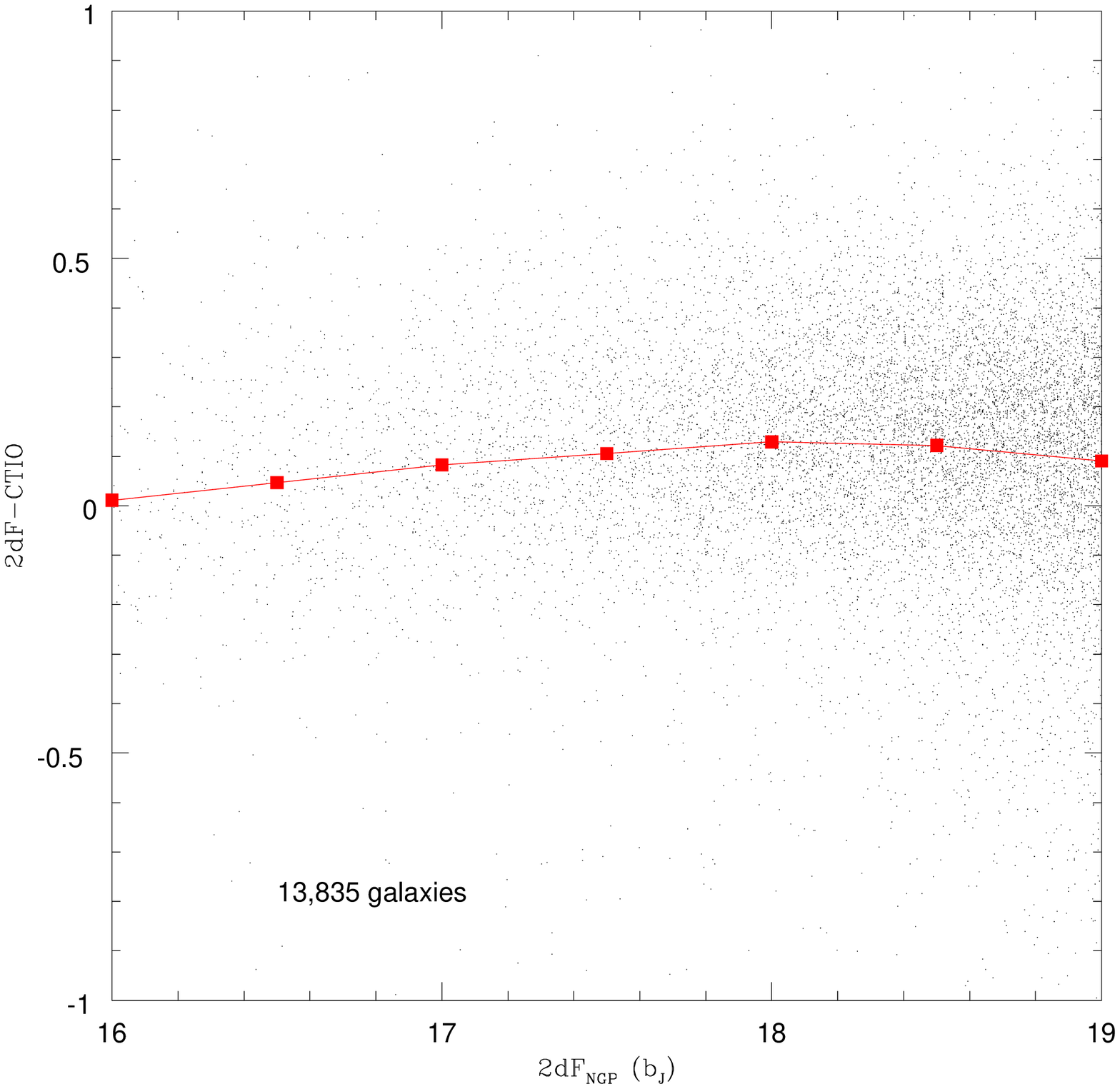}
\caption[\small{Comparison of our CTIO photometry with the 2dFGRS in the NGC.}]{This plot
shows 13,835 matching galaxies in the NGC from our CTIO survey and that of the 2dFGRS.
We have plotted the 2dF b$_J$ magnitude against the difference in magnitudes from the surveys,
2dF-CTIO, with the mean of this difference plotted in half magnitude bins.
We converted the CTIO data to the 2dF b$_J$ magnitude system using both our B Landolt and
R Kron-Cousins band-passes in conjunction with the colour equation b$_J$=B-0.17(B-R) from
Pimbblet et al. (2001).  One can see a clear zero-point and scale error over the magnitude
range shown with the zero-point difference peaking at 0.13 mag. when b$_J$=18.  For b$_J>16.0$
the mean correction is 0.1$\pm0.005$ mag.}
\label{fig:2dFphotngp}
\end{figure}

\begin{figure}
\centering
\includegraphics[width=3in,totalheight=3in]{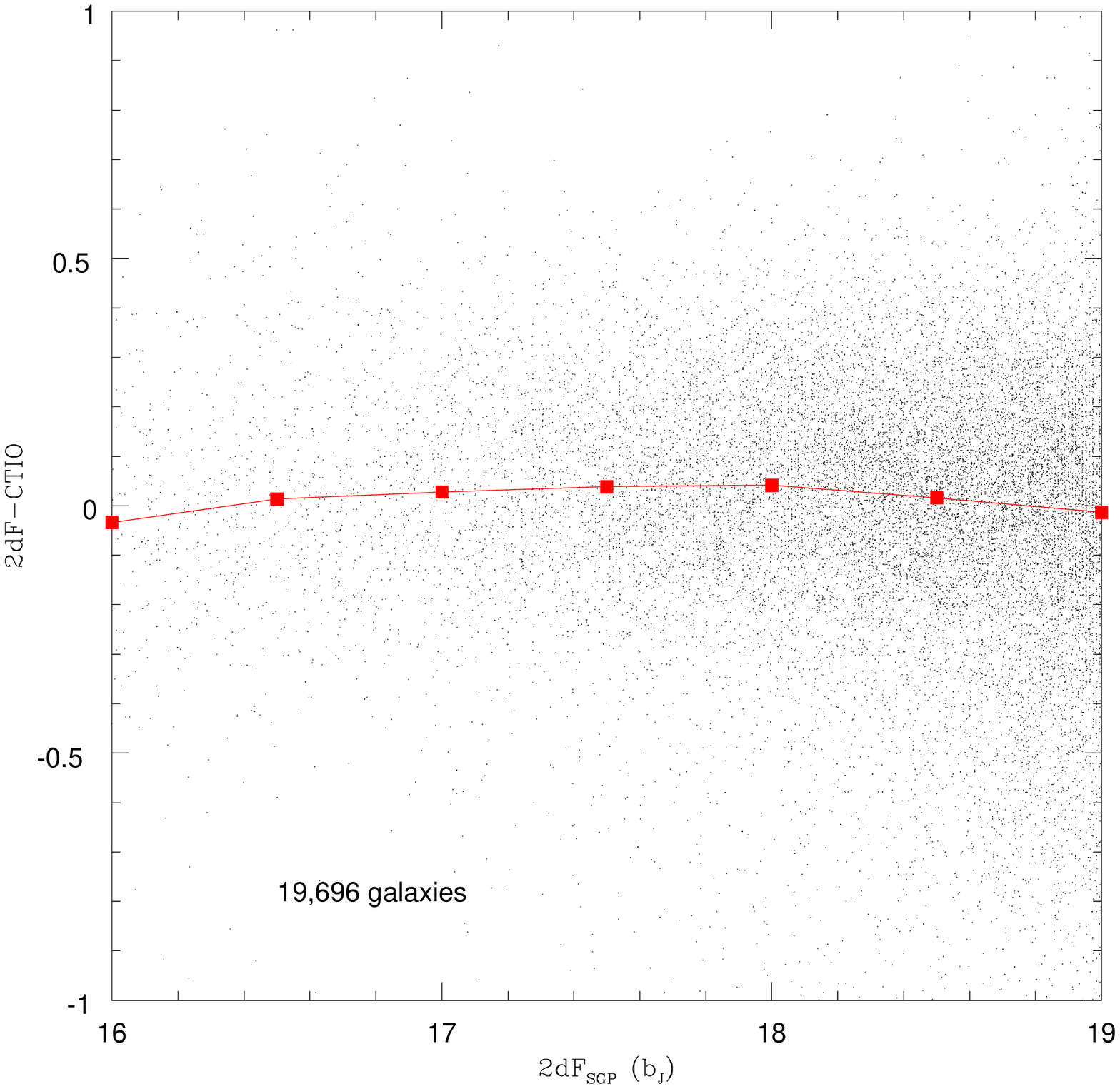}
\caption[\small{Comparison of our CTIO photometry with the 2dFGRS in the SGC.}]{This plot shows
19,696 matching galaxies in the SGC from our CTIO survey and that of the 2dFGRS.  We have plotted
the 2dF b$_J$ magnitude against the difference in magnitudes from the surveys, 2dF-CTIO, with
the mean of this difference plotted in half magnitude bins.  We converted the CTIO B-band data
to the 2dF b$_J$ magnitude system using both our B Landolt and R Kron-Cousins band-passes in
conjunction with the colour equation b$_J$=B-0.17(B-R) from Pimbblet et al. (2001).  The
agreement in the SGC is good with the mean of the magnitude difference $<0.05^+_-0.004$ mag.
over the whole magnitude range shown.}
\label{fig:2dFphotsgp}
\end{figure}

As described in Section 2, our 297 deg$^2$ SGC field is a complete subset of the
southern 2dFGRS region.  Therefore, providing our photometry is consistent with 
that of the 2dF, their SGC n(z) distribution can help us understand the exact
nature of the apparent under-density observed relative to the galaxy count model
of Metcalfe et al. (2001).   Furthermore, for b$_J>$17, the 2dFGRS photometry has
a zero-point that has changed slightly from the original APM data due to
re-calibration with external CCD magnitudes.  In the SGC this zero-point
difference is very small, APM-2dFGRS=0.02 mag., and so any comparisons of our CTIO
data with the 2dFGRS magnitudes will also be good checks of the original APM
photometry for b$_J>17$.  In the NGC this zero-point difference is
APM-2dFGRS=0.08 mag. (Cole priv. com.).  In order to check how our photometry compares
with that of the 2dFGRS we have matched all galaxies from the two surveys using matches
within a 3.5$''$ radius.  As in Section 4.1.1, we use both our B Landolt and R
Kron-Cousins galaxy photometry in conjunction with the colour equation from
Pimbblet et al. (2001) in order to convert to the 2dFGRS b$_J$ filter.  Figs.
\ref{fig:2dFphotngp} and  \ref{fig:2dFphotsgp} show these comparisons for the NGC
and SGC regions respectively.  Note that in the NGC only $\sim$2/3 of our fields
overlap with those of the 2dFGRS as opposed to 100\% in the SGC explaining the
factor of $\sim$1.5  difference in the  number of matching galaxies.
We have plotted the 2dF b$_J$ magnitude against the magnitude difference  
2dF-CTIO for each field with the mean of the residual illustrated by the squares
in each half magnitude bin.  The 2dF galaxy photometry is of course based on that
of the APM Galaxy Survey (Maddox et al. 1990a) with rms errors on the source
magnitudes of 0.2-0.25 mag., and this can be seen via the large scatter around the
mean in each magnitude bin.  We have shown in Section 3.1 that our CCD photometry
is much more accurate than this with rms errors of 0.1 mag. at B=18 and 0.05 mag.
at B=16, and therefore we expect any zero-point and scale errors  between the two
surveys to be contained within the photographic 2dF data.  The NGC comparison in
Fig. \ref{fig:2dFphotngp} shows clear evidence of both a zero-point and scale 
error.  We find  good agreement at b$_J$=16 but the scale error induces a maximum
zero-point error of 0.13 mag. at B=18 with a mean correction of 0.1 mag. for b$_J>$16.
For the SGC field in Fig. \ref{fig:2dFphotsgp} we find a much better agreement
with no appreciable scale error and the mean of the residuals agreeing to within
0.05 mag. over the whole magnitude range shown.  

\section{CTIO Number Counts}

\begin{figure}
\centering  
\includegraphics[width=3in,totalheight=3in]{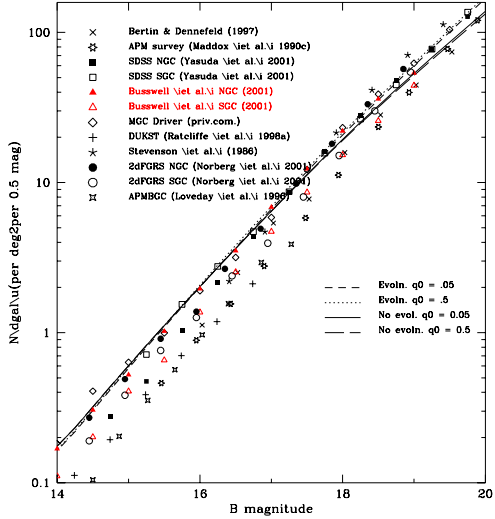}
\caption[\small{Our CTIO B-band galaxy number counts along with other data from
the literature.}]{Our B band galaxy number counts.  The filled and unfilled
triangles show our number counts in the North and South Galactic Caps
respectively.  The filled and unfilled  squares show the SDSS Commissioning Data
in the Northern and Southern equatorial strips (Yasuda et al. 2001), the circles
the 2dF Galaxy Redshift Survey (2dFGRS) data, the diamonds the Millenium Galaxy  
Catalogue (Driver priv. com.) results, the x's the data of Bertin \& Dennefeld  
(1997) and  the star-like symbols show the data of Stevenson et al. (1986). 
The APM counts (Maddox et al. 1990c) are indicated by the asterisk symbol and include the 
faint-end photometry correction of Metcalfe et al. (1995).  The crosses show the Durham/UKST data of  
Ratcliffe et al. (1998) and the filled diamonds the APM bright galaxy data (for
B$<$16.6 - for B$>$16.6 the APM-Stromlo counts are shown by the filled diamonds)
of Loveday et al. (1996).  The curves are Pure Luminosity Evolution (PLE) models
of Metcalfe et al. (2001) and they are explained further in section 4.3.   See  
text for details of dust corrections.  }
\label{fig:counts_b}
\end{figure}

\begin{figure}
\centering
\includegraphics[width=3in,totalheight=3in]{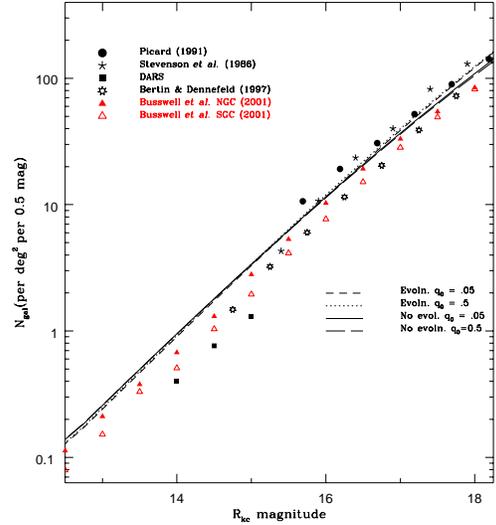}
\caption[\small{Our CTIO R-band galaxy number counts along with other data from
the literature.}]{Our R band galaxy number counts.  The filled and unfilled
triangles show our number counts in the North and South Galactic Caps
respectively.  The filled circles show the data of Picard (1991), the skeleton
stars that of Stevenson et al.(1986), the filled squares are the DARS data
(Metcalfe et al. 1998) and the open stars show the data of Bertin \& Dennefeld
(1997).  The curves are Pure Luminosity Evolution (PLE) models of Metcalfe et al.
(2001) which assume the presence of dust in late type spiral galaxies.  The
models are explained further in section 4.3. See text for details of dust
extinction corrections.}
\label{fig:counts_r}
\end{figure}

\begin{figure}
\centering
\includegraphics[width=3in,totalheight=3in]{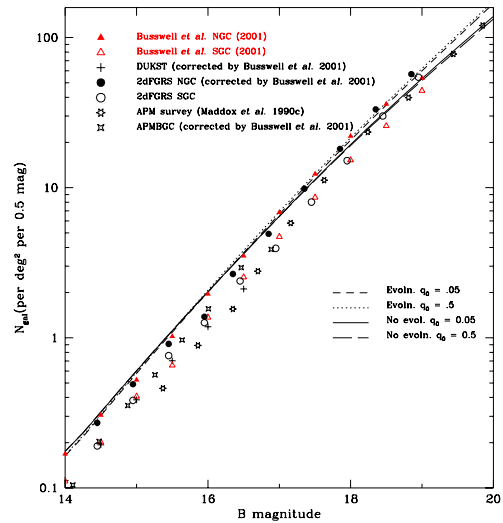}
\caption[\small{Our CTIO B-band galaxy number counts along with other corrected data.}]{Here we show
a similar plot to Fig. \ref{fig:counts_b}, but we impose the 0.313, 0.24, and 0.1 mag. photometry
corrections derived for the APMBGC, DUKST and 2dF NGC data (for b$_J>$16.0) respectively.
There is also a small 0.08 mag dust correction applied to the APM and APMBGC data (see text for
details). }
\label{fig:counts_rev}
\end{figure}

Figs \ref{fig:counts_b} and \ref{fig:counts_r} show our CTIO galaxy number
counts in the B and R bands respectively, which are both corrected for galactic
extinction using the Schlegel et al. (1998) dust maps.  Our B-band number counts
in the NGC agree very well, firstly with the SDSS data in the magnitude range  
$16.5<$B$<18.5$ and also with the MGC for $15.5<$B$<18.5$.  It should be noted
that all three data sets were taken in overlapping regions of sky - a 32 and 90
deg$^2$ overlap with the MGC and SDSS data respectively in the NGC.  Our CTIO   
data also overlaps with 163 deg$^2$ of the 2dFGRS in the NGC and 297 deg$^2$ in 
the SGC.  Our SGC number counts agree extremely well with the 2dFGRS, which is
important given the large overlap, but our NGC data is showing significantly more
galaxies than that of the 2dF team.  This could be due to real differences in the
galaxy number density  between the current 740 deg$^2$ 2dF field and our 297     
deg$^2$ field, or due to zero-point/scale differences in the galaxy photometry.

The shape and normalisation of our NGC number counts also agrees
remarkably  well with the no-evolution predictions of the low-$q_0$ model (see
section 4.3 for model details) from B=18 to as bright as B=14.  The shape of our
B-band number counts in the SGC agrees well with our NGC data set, but the
striking difference is that of the normalisation, which is calculated in each 0.5
magnitude interval in Table \ref{table:countsb}. We a find an average 30.7\%
deficiency in galaxy numbers for our southern data compared to the north in the
magnitude range 14$<$B$<$18.  At B=19 this normalisation discrepancy has dropped
to 17.3\% , although the galaxy incompleteness is a
factor in this magnitude bin.  The 2dF number
counts also show a north-south difference, but only 20\% in galaxy numbers at
B=18, with this difference virtually zero at B=19.  The survey area of our SGC
data was a subset of the 4300 deg$^2$ sky covered by the APM galaxy catalogue and
our SGC number counts agree very well with the Metcalfe et al. (1995) corrected APM counts despite the
order of magnitude less of sky coverage on our part.  Our SGC number counts also
agree well with the CCD data of Bertin \& Dennefeld (1997), who cover 62 and 83
square degrees at high declinations ($|\delta|>40^{\circ}$ in the North and South
Galactic Caps respectively.  Our CTIO counts, the 2dFGRS and MGC counts and those
of Yasuda et al. have all been corrected for dust using the Schlegel et al.
(1998) extinction maps.  The APM and DUKST counts and those of Bertin \&
Dennefeld are not corrected for dust and the reason for this was that the vast
majority of all these observations were in the SGP where it was orginally thought
dust extinction was negligible.  In fact Schlegel et al. predict values of 0.03 mag.
in the B-band at the SGP and an average of 0.08 mag. over the whole SGC.
           
Our R-band galaxy counts show a similar trend in the normalisation
difference, where the number discrepancies of the SGC counts relative to those in
the SGC are tabulated in Table \ref{table:countsr}.  The average percentage
discrepancy is 22.2\% in the 13$<$R$<$17 range, and again this normalisation
difference drops for fainter magnitudes.  At R=18 there is only a 2.5\%
discrepancy in the galaxy numbers, although galaxy incompleteness is again a
factor.  The shape of both the data sets in the
13$<$R$<$17 range is consistent with the no-evolution models, but there is a
slight normalisation discrepancy in that all the data sets show less galaxies
compared to the models for R$<15.5$.  There is a noticeable dip in both our
number counts compared to the models between about R=13.5 and R=15.5 and there is
good agreement with Bertin \& Dennefeld, as in the B filter.  Our data lies
substantially below that of Picard et al. (1991), who cover 386 square degrees in
both the NGC and SGC.   In Fig. \ref{fig:counts_b} our CTIO number counts have  
been corrected for dust using the Schlegel et al. (1998) extinction maps but none
of the other data sets were de-reddened when originally published.  This is
because, either it was originally thought that a particular data-set suffered
from negligible extinction, as in the case of the Stevenson points in the SGC, or
the reddening correction was not known due to the absence of any accurate dust  
maps at the time of publication eg. Picard (1991).

\begin{table}
\centering 
\begin{tabular}{||c|c|c|c||} \hline
               B   &  N$_{gal}^{NGC}$       &  N$_{gal}^{SGC}$       &  N-S \\
                   &                        &                        &  discrepancy \\ \hline
           12.25-12.75 &  0.031 &  0.017 &  45.2    \\
           12.75-13.25 &  0.047 &  0.024 &  48.9    \\
           13.25-13.75 &  0.066 &  0.067 &  -1.5    \\
           13.75-14.25 &  0.168 &  0.111 &  33.9    \\
           14.25-14.75 &  0.305 &  0.201 &  34.1    \\
           14.75-15.25 &  0.524 &  0.406 &  22.5    \\
           15.25-15.75 &  1.021 &  0.655 &  35.8    \\
           15.75-16.25 &  1.959 &  1.367 &  30.2    \\
           16.25-16.75 &  3.508 &  2.535 &  27.7    \\
           16.75-17.25 &  6.855 &  4.714 &  31.2    \\
           17.25-17.75 & 12.353 &  8.619 &  30.2    \\
           17.75-18.25 & 22.059 & 15.283 &  30.7    \\
           18.25-18.75 & 36.047 & 25.779 &  28.5    \\
           18.75-19.25 & 53.604 & 44.353 &  17.3    \\
           19.25-19.75 & 59.521 & 76.012 & -27.7    \\  \hline
\end{tabular}
\caption[\small{Our tabulated B-band number counts in the NGC and SGC.}]{\small{Here we
show our dust-corrected B-band number counts per degree squared per half magnitude for our NGC and
SGC fields.  We have also shown the number discrepancy (as a percentage) of the SGC data
relative to the NGC data. } }
\label{table:countsb}
\end{table}

\begin{table}
\centering
\begin{tabular}{||c|c|c|c||} \hline
               R  &  N$_{gal}^{NGC}$       &  N$_{gal}^{SGC}$       &  N-S \\
                  &                        &                        &  discrepancy \\  \hline
           11.25-11.75 &   0.019 &   0.020 &  -0.5    \\
           11.75-12.25 &   0.094 &   0.043 &  54.3    \\
           12.25-12.75 &   0.113 &   0.079 &  30.1    \\
           12.75-13.25 &   0.211 &   0.152 &  28.0    \\
           13.25-13.75 &   0.379 &   0.331 &  12.7    \\
           13.75-14.25 &   0.676 &   0.509 &  24.7    \\
           14.25-14.75 &   1.309 &   1.036 &  20.9    \\
           14.75-15.25 &   2.810 &   1.956 &  30.4    \\
           15.25-15.75 &   5.315 &   4.136 &  22.2    \\
           15.75-16.25 &  10.294 &   7.683 &  25.4    \\
           16.25-16.75 &  19.157 &  15.078 &  21.3    \\
           16.75-17.25 &  33.120 &  28.317 &  14.5    \\
           17.25-17.75 &  54.423 &  49.305 &   9.4    \\
           17.75-18.25 &  84.271 &  82.154 &   2.5    \\
           18.25-18.75 & 113.543 & 124.624 &  -9.8    \\  \hline
\end{tabular}
\caption[\small{Our tabulated R-band number counts in the NGC and SGC.}]{\small{Here we show
our dust-corrected R-band number counts per degree squared per half magnitude for our NGC and SGC fields.
We have also shown the number discrepancy (as a percentage) of the SGC data relative to the
NGC data. }}
\label{table:countsr}
\end{table}

Fig. \ref{fig:counts_rev} shows selected number count data with the   
appropriate zero-point corrections of 0.10 mag. to the 2dFGRS NGC data (for 
b$_J>$16.0), 0.24 mag. to that of the DUKST and 0.31 mag. to the APMBGC data.
The DUKST corrected data is now significantly altered, but the counts are still
much lower than the q$_0$=0.05 evolution model.  We have also made a small dust correction to the APM
and APMBGC data shown in Fig. \ref{fig:counts_rev}.  A standard
A$_B$=C(cosec($b$)-1) extinction law was originally assumed by Maddox et al. for
the APM photometry, which corresponds to A$_B$=0 at the poles and a maximum
A$_B\sim$0.03 mag. at $b=50^{\circ}$, averaging $\sim0.01$ mag over the whole APM
area.  The Schlegel et al. (1998) dust maps, which are used to correct the 2dF
counts and our own CTIO data, predict 0.02-0.03 mag. of extinction, even at the
poles, and our average CTIO dust correction was 0.08 mag. in the SGC.  We have
therefore corrected the APM counts and the APMBGC data by an additional 0.08 mag
(so the total correction is 0.08 mag. for the APM counts and 0.39 mag. for the
APMBGC data).  The original APM counts now show no galaxy number deficiency for
B$>$18, but still indicates a large local hole at brighter magnitudes over a huge
4300 deg$^2$ area.

The 2dF data shown in Fig \ref{fig:counts_rev} shows excellent
agreement with our CTIO data in both galactic caps.  The 0.1 mag. zero-point
correction we have applied to the 2dF NGC data now means that their north-south
difference agrees with ours almost exactly for B$<$18.  This agreement with our
data in each galactic cap is interesting given the larger angular areas of 740
deg$^2$ and 1094 deg$^2$ of the 2dFGRS NGC and SGC fields respectively.
Therefore it would seem that our survey areas of 242 deg$^2$ and 297 deg$^2$ are
fairly typical samples of the galaxy distribution in these regions of the
Universe.

\section{Number Redshift Distributions}

\subsection{The Durham/UKST N(z) Distribution}

\begin{figure}
\centering
\includegraphics[width=3in,totalheight=3in]{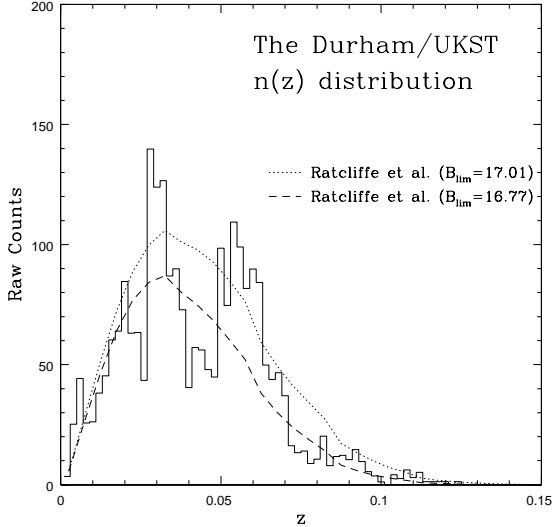}
\caption[\small{The DUKST n(z) distribution compared to predictions from the Ratcliffe et al. luminosity
function.}]{The plot shows the DUKST n(z) distribution, which has been constructed using the ``best sample''
from Ratcliffe et al. (1998) with the corresponding magnitude limits and completeness corrections for the raw
EDSGC photometry in all 60 fields. The
average magnitude limit of the 60 fields, taking into account the photometry correction applied by Ratcliffe
et al. is b$_J$=16.86-0.05=16.81. The dotted line shows the prediction using the original luminosity function
of Ratcliffe et al. (1998a) calculated
from the data itself using the appropriate B Landolt magnitude limit of 16.86-0.05+0.20=17.01.  The dashed line
shows a prediction except with the 0.24 mag. photometry correction applied to the magnitude limit, giving
B$_{lim}=17.01-0.24=16.77$.}
\label{fig:DUKSTnza}
\end{figure}

The zero-point discrepancy between the DUKST photographic plate photometry and  
that of our CCD based data has important consequences in terms of the DUKST
number counts and n(z) distributions.   We have used the ``best sample'' from
Ratcliffe et al. (1998) using the uncorrected EDSGC photometry, taking into
account the magnitude limits and completeness corrections of each of the 60
fields, in order to re-construct the DUKST n(z) distribution shown in Fig.
\ref{fig:DUKSTnza}. The average magnitude limit for the DUKST ``best sample'' 
over all 60 fields is b$_J$=16.86 and the dotted line shows the n(z) predicted   
using the Ratcliffe et al. (1998a) luminosity function parameters. This has 
been calculated using Ratcliffe's ``best sample'' magnitude limit, first
brightened  by 0.05 mag. to take account of the DUKST photometry relative to that of
the EDSGC at b$_J$=19.5, and secondly 0.2 mag is added in order to convert to the
B Landolt system. We therefore use $B_{lim}=16.86-0.05+0.2=17.01$. In fact, 
this curve is equivalent to correcting all of the DUKST magnitudes by 0.24 mag.
\textit{and} the M$^{\star}$ value in the luminosity function by the same amount
as the two effects cancel each other out.   The dashed line shows a prediction
with the 0.24 mag. photometry correction applied to the magnitude limit,
giving B$_{lim}=17.01-0.24=16.77$, but with no alterations to the original value
of the M$^{\star}$ of Ratcliffe et al. (1998b).

\begin{figure}
\centering
\includegraphics[width=3in,totalheight=3in]{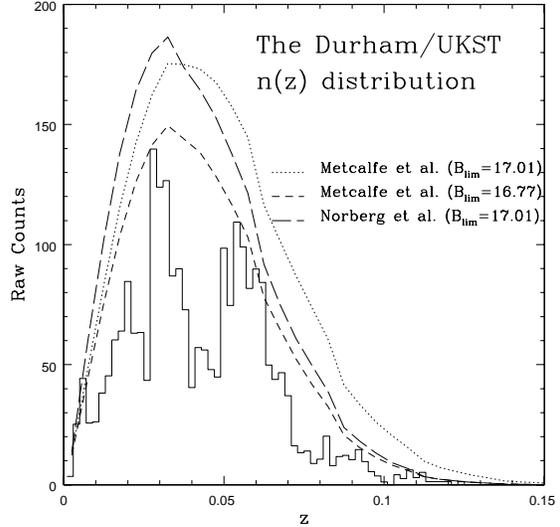}
\caption[\small{The DUKST n(z) distribution compared to predictions from other luminosity functions.}]{The
DUKST n(z) distribution compared to predictions from other luminosity functions. The dotted line shows a
prediction using the luminosity function of Metcalfe et al. (2001) calculated using the appropriate B Landolt
magnitude limit of 16.86-0.05+0.20=17.01. The
long dashed line shows a prediction from the new 2dFGRS luminosity function using the same magnitude limit (see
text for more details of this luminosity function).  Finally, the short dashed curve again shows a prediction using
the luminosity function
of Metcalfe et al. (2001) but calculated for the magnitude limit B$_{lim}$=16.77, which takes into account the 0.24
zero-point discrepancy found between our CTIO CCD data and the DUKST photographic photometry.}
\label{fig:DUKSTnzb}
\end{figure}

Fig. \ref{fig:DUKSTnzb} again shows the DUKST n(z) distribution but this   
time with predictions using the luminosity functions of Metcalfe et al. (2001)   
and Norberg et al. (2001).  The b$_J$ luminosity function of Norberg et al. (long
dashed curve) has been calculated using 110,500 galaxies from the 2dFGRS at z=0, 
taking account of evolution, the distribution of magnitude measurement errors and
small corrections for incompleteness within the 2dF catalogue.  We have used
their published values of M$_{b_J}^{\star}$=-19.66 (M$_{B}^{\star}$=-19.46) and
$\alpha=-1.21$.  A $\phi^{\star}$=2.6x10$^{-2}$Mpc$^{-3}$ was chosen so that the
predicted galaxy number count from the Norberg et al. luminosity function matched
that of the Metcalfe et al. model at B=19.5.  
We have also shown a prediction from the
luminosity function of Metcalfe et al. (2001) using a magnitude limit of
B$_{lim}$=17.01 (dotted curve) and the CTIO corrected magnitude limit of   
B$_{lim}$=16.77 (short dashed curve).

The predictions from the two luminosity functions agree reasonably well
for z$<$0.05 but the Metcalfe et al. model predicts slightly more galaxies at
higher redshift.  However, both models clearly over-predict the DUKST n(z) at all
redshifts.  This clear deficiency of galaxies was originally considered
surprising by Ratcliffe et al. given the DUKST covers 1500 deg$^2$ but it was
thought that, because of the relatively bright magnitude limit of b$_J$=16.86 and  
shallow redshift depths, that significant large scale structure at low redshift  
could explain this.  By taking into account the zero-point correction of 0.240   
and using the Metcalfe et al. model with B$_{lim}=$17.01-0.240=16.77, we find
that this apparent underdensity of galaxies is not as large as first thought.
Even so the corrected Metcalfe et al. model still over-predicts the n(z)
distribution at virtually all redshifts and so we draw a similar  conclusion to
that of Ratcliffe et al. in that, even after our photometry correction is
applied, significant large scale structure is still observed in the DUKST n(z)
distribution.  The dashed line in Fig \ref{fig:DUKSTnzb} shows this
model which indicates that the deficiency of galaxies relative to the Metcalfe et al.
model is still apparent until at least z=0.1  Relative to this 
photometry-corrected Metcalfe et al. model the DUKST n(z) distribution shows three clear holes
in the galaxy distribution in the redshift ranges 0.005$<$z$<$0.025,
0.03$<$z$<$0.055 and 0.06$<$z$<$0.09 with number discrepancy
percentages of $\sim$40\%, 45\% and 50\% 
respectively. In Section 6 we will analyse these apparent under-densities in
more detail in conjunction with the results from the 2dFGRS n(z) distribution
which we discuss in Section 5.

\subsection{The 2dFGRS N(z) Distributions}

The good photometry agreement in the SGC field between us and the 2dFGRS means
that we are able to use the 2dF SGC n(z) distribution in conjunction with the
galaxy count model of Metcalfe et al. (2001) described in 4.3, to perform a
consistent analysis of the observed deficiency of galaxies in the SGC as a
function of redshift.  We can also make the appropriate zero-point correction in
the NGC so we are then consistent with the 2dF photometry in order to investigate
the 2dF NGC n(z) distribution.

\begin{figure}
\centering
\includegraphics[width=3in,totalheight=3in]{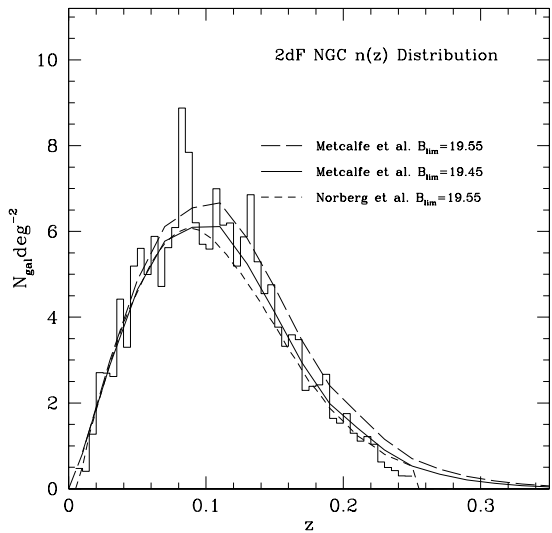}
\caption[\small{The 2dFGRS NGC n(z) distribution.}]{Here we show the n(z) distribution from the 2dFGRS in their NGC
field along with two galaxy number count predictions using the luminosity function of Metcalfe et al. (2001) (solid
and long dashed lines).
We also show a prediction using the 2dF luminosity function (Norberg et al. 2001) (short dashed line).   The mean
magnitude limit of the APM-based photographic plates in the NGC is b$_J$=19.35 and so, assuming B-b$_J$=0.20 mag, we
adopt a magnitude limit, m$_{lim}$ of B=19.55 using each of the Norberg et al. and Metcalfe et al. models.  The 
second
prediction of the
Metcalfe et al. model uses a magnitude limit corrected brighter by 0.1 mag. to B=19.45 in accordance with the 
zero-point
difference we found
between the 2dF NGC photometry and that of our CTIO data.  }
\label{fig:2dFnzngp}
\end{figure}

\begin{figure}
\centering
\includegraphics[width=3in,totalheight=3in]{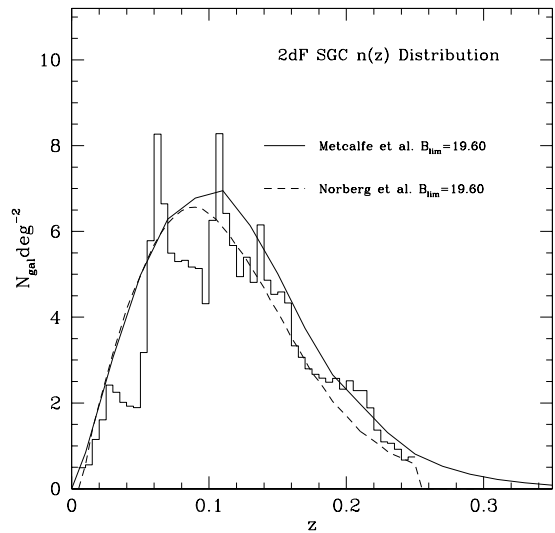}
\caption[\small{The 2dFGRS SGC n(z) distribution.}]{Here we show the n(z) distribution from the 2dFGRS in their SGC
field and predictions using the luminosity functions of Metcalfe et al. (2001) (solid line) and Norberg et al. (2001)
(dashed line).  There
are two clear ``holes'' in the 2dF SGC galaxy distribution in the ranges 0.03$<$z$<$0.06, with an under-density of
$\sim$35-40\%, and 0.07$<$z$<$0.1 where the galaxy density deficiency is $\sim$25\%.
The mean magnitude limit of the photographic plates in the SGC is b$_J$=19.40 and so, assuming B-b$_J$=0.20, we
adopted a magnitude limit, m$_{lim}$ of B=19.60 when computing the n(z) predictions.}
\label{fig:2dFnzsgp}
\end{figure}

We show the 2dFGRS n(z) distributions in Figs. \ref{fig:2dFnzngp} and
\ref{fig:2dFnzsgp} obtained from the 100k data release (Colless et al. 2001).  The
mean magnitude limit of the 2dF photographic plates is b$_J$=19.35 and
b$_J$=19.40 in the NGC and SGC respectively. Due to uncertainties in the 
effective areas and completeness, we have normalised the data such that the ratio of the
predicted to the observed number of galaxies matches the corresponding ratio in the 
n(B) counts to the same magnitude limits. In Fig. \ref{fig:2dFnzngp}
we have plotted three model curves, two of which use the Metcalfe et al. (2001)
luminosity function (long dashed and solid curves) and a third which uses the
luminosity function of Norberg et al. (2001) (short dashed curve) described in
section 3.4.2.  The long dashed and short dashed curves were both calculated
using the appropriate mean magnitude limit, B$_{lim}$=19.55, of the APM-based
plates in the NGC (assuming B-b$_J$=0.2).  We only show the Norberg et al. model
at B$_{lim}$=19.55, since any photometry correction would be also applied to the
Norberg et al. M$^{\star}$; this would leave the n(z) virtually unchanged.

The models agree at low redshift but the B$_{lim}$=19.55 Metcalfe et al. model
significantly over-predicts the data, particularly at the higher redshifts
plotted.  The solid curve, however, takes into account our zero-point correction
derived in the previous section, equal to 0.1 mag. for b$_J>$16, and we therefore use
a magnitude limit, B$_{lim}$=19.55-0.1=19.45.  This model now shows excellent
agreement with the observed NGC data and it is interesting that this zero-point
correction is vital in order that the Metcalfe et al. model does not over-predict
the data.  This excellent agreement of the data with the Metcalfe et al. and
Norberg et al. models at all redshifts, indicates that the galaxy distribution
appears to be fairly uniform in the 2dF NGC field with no evidence for any huge  
under-densities as we saw for the DUKST field.

Fig. \ref{fig:2dFnzsgp} shows the 2dF SGC n(z) distribution along with
two model predictions using the Metcalfe et al. luminosity function (solid curve)
and that of Norberg et al. (dashed curve).  Our photometry comparison with the
2dF SGC data in the previous section showed up no significant zero-point error
and so we used the appropriate magnitude limit of B$_{lim}$=19.60 for each model.
The main result of these model comparisons so far has been that the Metcalfe et
al. model predicts more galaxies at higher redshift and this case is no
different, but the Metcalfe et al. model shows a slightly better agreement with
the data than the Norberg et al. model which under-predicts the data for
z$>$0.14.  The most striking feature of the plot however is the two clear
``holes'' in the galaxy distribution in the ranges 0.03$<$z$<$0.055, with an
under-density of $\sim$35-40\%, and 0.06$<$z$<$0.1 where the density deficiency
is $\sim$25\%. We now analyse these underdensities in the galaxy distribution in
more detail. 

\section{Discussion}

\begin{figure}
\centering
\includegraphics[width=3in,totalheight=3in]{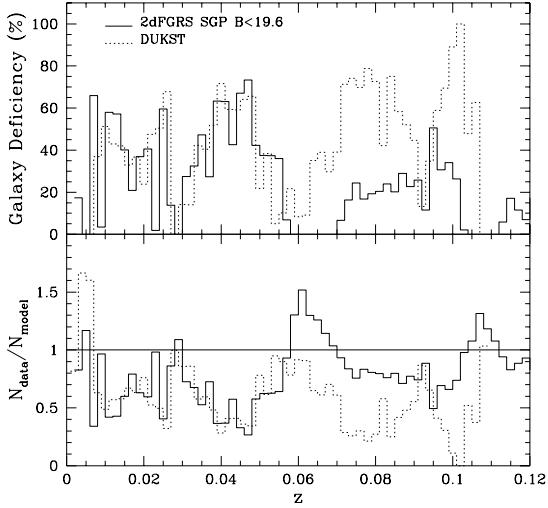}
\caption[\small{A comparison of the  number deficiency in the DUKST and 2dF SGC n(z) distributions.}]{In 
the upper panel we show the galaxy number deficiency in the DUKST and 2dF SGC for B$_{lim}$=19.6 
n(z) distributions plotted as a function of redshift.  
We have calculated this using the histograms in Figs. \ref{fig:DUKSTnzb} and \ref{fig:2dFnzsgp} 
relative to the appropriate prediction using the Metcalfe et al. (2001) luminosity function.  We define the 
galaxy deficiency (in \%), 
D$_g$(z)=100(N$_{model}$(z)-N$_{data}$(z))/N$_{model}(z)$ where N$_{data}$ is the number of galaxies from 
the appropriate survey data and N$_{model}$ is the number predicted from the model of Metcalfe et al.  
The lower panel illustrates the ratio of the quantity N$_{data}$/N$_{model}$ for the DUKST and 2dF SGP fields, 
also as a function of redshift.  }
\label{fig:bothholes}
\end{figure}

We have attempted to analyse the depth and angular size of the apparent ``hole''
in the SGC galaxy distribution using data from the APM, 2dFGRS, DUKST and 2MASS
surveys.  One of the major discoveries of this paper has been the
large galaxy-number deficiency relative to the Metcalfe et al. model in the
DUKST and 2dF n(z) distributions.  In fact there are distinct similarities
between the two n(z) distributions which we attempt to quantify in Fig.
\ref{fig:bothholes}.  There are two panels in these histogram  plots where the
upper panel shows galaxy number deficiency vs redshift and the lower panel shows
N$_{gal}$/N$_{model}$ again vs redshift.  We define the percentage galaxy number deficiency
as:

\begin{equation}
D_g(z)=100\frac{(N_{model}(z)-N_{data}(z))}{N_{model}(z)}
\end{equation}

\noindent where N$_{model}$ is the number of galaxies predicted using the
Metcalfe et al. (2001) luminosity function and N$_{data}$ is the number of
galaxies from the appropriate survey data.  In each panel, predictions for the
DUKST survey are shown by the solid lines and those of the 2dFGRS by the dashed
line.  In the upper panel the four holes found in the DUKST n(z) distribution are
clearly illustrated by the four peaks in the solid histogram with the two peaks of the
dashed histogram showing the under-densities found in the 2dF SGC n(z).

The similarity between the 2dF and DUKST n(z) distributions is quite
striking. The two  under-densities seen in the 2dF SGC n(z) in the redshift   
ranges 0$<$z$<$0.055 and 0.06$<$z$<$0.1 are also clear features in the DUKST   
n(z).  In fact, the galaxy discrepancy in the range 0$<$z$<$0.055 for the two  
surveys is almost the same magnitude in size with the histograms showing a very 
similar shape.  The second under-density in the 2dF SGC n(z) is less pronounced 
than its DUKST counterpart, but they still cover very similar redshift ranges.  
The rise in the galaxy number density between the two 2dF SGC under-densities is
also seen in the case of the DUKST survey.  Given that the 2dF SGC field is
entirely contained within the areas of sky observed for the DUKST survey we claim
that the the similarities we have described in the two n(z) distributions are  
both artefacts of the \textit{same} features in the galaxy distribution.  Since
the DUKST observes a larger 1500 deg$^2$ region than the 2dFGRS (who have
redshifts for galaxies covering $\sim$300 deg$^2$ in the SGC at present) it could
be that there is significant large scale structure observed by DUKST that is not
seen by the 2dFGRS explaining the larger and more numerous galaxy number
discrepancies in the DUKST n(z) distribution.

\begin{figure}
\centering
\includegraphics[width=3in,totalheight=3in]{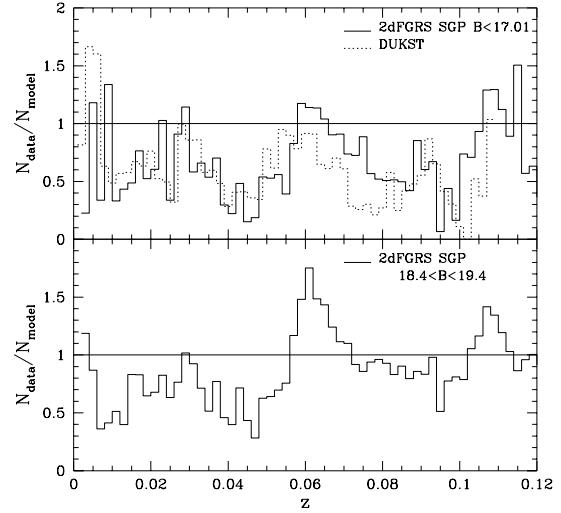}
\caption[\small{The 2dFSGC n(z) distribution with a bright magnitude limit imposed of B=17.01.}]{In the upper 
panel the histogram shows the the ratio of the quantity N$_{data}$/N$_{model}$ for the 2dFSGC n(z) distribution 
with an imposed magnitude limit of B=17.01, relative to the appropriate prediction using the the Metcalfe et al. 
(2001) luminosity function. The dotted histogram shows the DUKST n(z) as in Fig. \ref{fig:bothholes}. The lower 
panel shows the same ratio for the 2dFGRS for 18.4$<$B$<$19.4 relative to the Metcalfe et al. model with the same 
limits. }
\label{fig:biasplot}
\end{figure}

But there is evidence for an alternative explanation. For
z$>$0.06 shown in the lower panel of Fig. \ref{fig:bothholes}, the DUKST
survey shows significantly larger galaxy number discrepancies than the 2dFSGC n(z) relative to the
Metcalfe et al. model. It is possible that galaxies can be biased tracers of the overall mass
distribution and that intrinsically more luminous galaxies are then predicted to show
stronger clustering properties, with regions of very high and low galaxy number 
densities.   We suspect that the SGC might be under-dense and therefore the fact
that the DUKST samples intrinsically brighter galaxies than the 2dF over all
redshifts shown in Fig. \ref{fig:bothholes}, may mean that we are seeing the
effects of bias.

Since this clearly needs further investigation, we have examined how the galaxy distribution
varies with apparent magnitude. The results are shown in Fig. \ref{fig:biasplot}. 
In the upper panel we have plotted N$_{data}$/N$_{model}$ for the 2dFGRS SGC data relative to the Metcalfe et al. 
(2001) model, each with an imposed magnitude limit of b$_J$=16.81 (B=17.01) to match the 
DUKST limiting magnitude. The DUKST n(z) is indicated as in Fig. \ref{fig:bothholes} 
by the dotted histogram for reference. It is very interesting that we now see a better agreement between the 2dF and 
DUKST n(z) for z$>$0.06. Similarly, in the lower panel we have imposed magnitude limits of 
18.4$<$B$<$19.4 on the 2dFGRS SGC data and the Metcalfe et al. prediction in order to examine whether the
galaxy number deficiencies seen in the bright data are compensated for by fainter galaxies. We can see from the 
lower panel of Fig. \ref{fig:biasplot} that there is no significant effect for z$<$0.06, while the galaxy distribution 
for z$>$0.06 indicates a significantly reduced galaxy deficiency. This is entirely due to the different galaxy 
selection criterion we have used via
the two magnitude limits of B$<$17.01 and 18.4$<$B$<$19.4, and suggests that the intrinsically 
brighter galaxies are clustered more strongly on \gsim50\mpc\  scales, despite the fact that 
the amplitude of the underdensities at z$<$0.06 remain the same; the galaxies
here are sub-L$^{\star}$ in both samples with approximately similar space densities, and
so even this result may be in accordance with the notion of bias.

\begin{figure}
\centering
\includegraphics[width=3in,totalheight=3in]{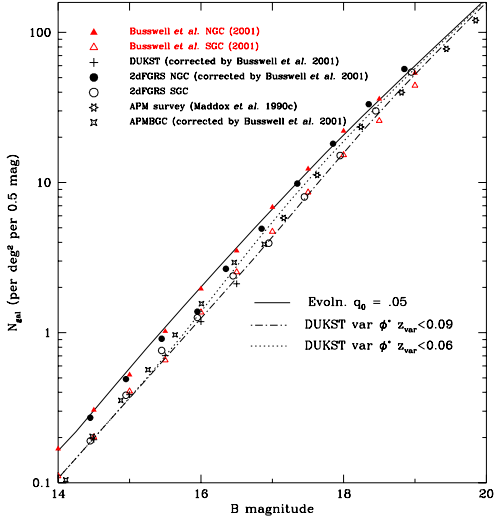}
\caption[\small{B-band galaxy number counts of survey data corrected using our photometry.}]{Here
we show a similar plot to Fig \ref{fig:counts_rev} with the 0.313, 0.24 and 0.1 photometry
corrections derived for the APMBGC, DUKST and 2dF NGC data respectively.
We have also plotted four variable $\phi^{\star}$ models shown by the short dashed,
dot-dashed, dotted and long-dashed curves, where the value of $\phi^{\star}$ is a function of
redshift (see text for detailed expanation).}
\label{fig:countsb_photcorr}
\end{figure}

Fig. \ref{fig:countsb_photcorr} shows our B-band galaxy number counts and our corrected APMBGC, 
2dF and DUKST counts (see Fig. \ref{fig:counts_rev}). For clarity, the bright original APM (B$<$17)
counts have been removed since we have no reliable photometry checks using our CTIO data here, and 
the faint-end Metcalfe et al. corrected APM counts are shown with our corrected APMBGC counts.    
These counts are shown with respect to two variable $\phi^{\star}$ models 
where the value of $\phi^{\star}$ is a function of redshift instead of the usual
constant value defined from the Metcalfe et al. (2001) B-band luminosity 
function. In these models the $\phi^{\star}$ is determined for
certain low redshift ranges (z$\le$z$_{var}$) by multiplying the usual Metcalfe et al. (2001) value by the factor   
N$_{data}$(z)/N$_{model}$(z) for the bin widths shown in Fig. \ref{fig:bothholes}. 
For z$>$z$_{var}$ the usual Metcalfe et al. (2001)   
$\phi^{\star}$ is used.  

The motivation for these variable $\phi^{\star}$ models is to locate the structure in the n(z)
which dominates the bright galaxy counts. 
Recall that the DUKST n(z) indicates three distinct deficiencies in the galaxy distribution below 
z$\approx$0.1. The DUKST z$_{var}<$0.09 and z$_{var}<$0.06 models 
were constructed then in order to see the effect on the predicted number counts, assuming that just the
first two and then all three of these holes have significant effects on the galaxy counts.  

The z$_{var}<$0.09 model fits the bright Southern CTIO counts well but slightly underpredicts the
APMBGC data. The z$_{var}<$0.06 overpredicts the Southern CTIO counts. Since
this redshift range appears to be unaffected by bias this indicates that the deficiency in the 
Southern CTIO counts is caused by structure to z$\lsim0.1$. 
The fact that the z$_{var}<$0.09 
model slightly underpredicts the APMBGC counts indicates that the galaxy deficiency of the hole may be less
over the whole SGC than the $\approx$30\% indicated in the counts from the DUKST, 2dFGRS SGC and 
Southern CTIO strips. A comparison of the APMBGC counts in Fig. \ref{fig:countsb_photcorr} with
the homogeneous model still indicates a 25\% deficiency over the full 4300 deg$^2$ of the APM 
Southern Survey. The conclusion from these variable $\phi^{\star}$ models then is that the  
Southern B-band counts are most consistent with there being a galaxy
number discrepancy for z\lsim0.1, which is of order $\approx$25\%.

These results agree very well with those of Frith et al. (2003), who investigated
the angular extent of the hypothesised deficiency of galaxies in the SGC using the publicly available 2MASS 
second incremental release data.  
The authors took $5^{\circ}$ declination
slices in the NGC and SGC with RA ranges similar to that of the 2dF regions, and plotted galaxy number counts in the 
K-band.   In total twelve
declination slices of data, each of $\sim300-400$ deg$^2$, were analysed totalling $\sim4300$ deg$^2$ from the two 
Galactic Caps. Since the APM, 2dF SGC and DUKST survey areas overlap with the three of the southern 2MASS 
strips, the work done by
Frith et al. is extremely relevant to the discussion in this paper.  Their conclusions also show strong evidence for a 
significant under-density of $\approx$30\%
in the galaxy number density in their SGC areas.  This agrees well with the 25\%
deficiency that we have shown in this paper is implied by the APM and APMBGC data.  The DUKST field then extends 
further south than is currently sampled by 2MASS,
to $\delta$=-43$^{\circ}$ and the APM area even further to $\delta$=-70$^{\circ}$. This then implies
a hole in the SGC galaxy distribution of 100$^{\circ}$x60$^{\circ}$, which extends to z$\approx$0.1.   
This corresponds to a huge volume of space of
$\approx$10$^7$~$h^{-3}$~Mpc$^3$ and implies significant power on large scales of $\sim$100-200\mpc.

To illustrate how excess power on larger scales increases the chances of finding
a 25\% galaxy deficiency over these volume sizes, we have used
the 3-D analogue of equation 45.6 in Peebles (1980).  By assuming a power law
form of the spatial two-point correlation function out to a given scale
length, we can calculate the probability of finding a given number deficiency of
galaxies over a volume of space defined by a sphere of radius 150$h^{-1}$Mpc.  In
this simplified scenario, the fluctuation in galaxy number over the expected  
galaxy number can be written as:

\begin{equation}
\frac{<(N-\bar{N})^2>^{\frac{1}{2}}}{\bar{N}}=\frac{\left(1+\frac{4\pi}{V}\int_0^{r_{s}}
\xi(r)r^{2}dr\right)^{\frac{1}{2}}}{\bar{N}^{\frac{1}{2}}}
\end{equation}

\noindent where $\xi(r)$ is the two-point spatial function and V is the volume of
the sphere over which a particular survey is sampling.  If we write
$\frac{<(N-\bar{N})^2>^{\frac{1}{2}}}{\bar{N}}$=$\frac{\delta N}{\bar{N}}$ then:

\begin{equation}
\frac{\delta N}{\bar{N}}=\left(\frac{1}{\bar{N}}+\frac{3}{r_s^3}\left[\frac{r_0^{1.8}r_{cut}^{1.2}}{1.2}
\right]\right)^{\frac{1}{2}}
\end{equation}

\noindent where we assume the galaxy correlation length in proper coordinates,   
r$_0$=5.0$h^{-1}$Mpc, r$_{cut}$ in Mpc is the length scale to which we assume the
power law form of the correlation function extends, r$_s$=150$h^{-1}$Mpc and  
is the radius of the sphere defining our volume, $\bar{N}$ is the expected number
of galaxies in this volume and $\delta N$/$\bar{N}$ is the expected fluctuation
in the galaxy number.   We assume two cases, each corresponding to a different
value of r$_{cut}$.  In the first case r$_{cut}$=10$h^{-1}$Mpc.  If we set
$\bar{N}$=$(4\pi r_s^3\bar{n})/3$, where we assume $\bar{n}$=0.01 which is the   
mean galaxy density in units of $h^{3}$Mpc$^{-3}$, then we find that $\delta  
N$/$\bar{N}$=0.016.  This corresponds to an expected galaxy number fluctuation
over our sphere of radius 150$h^{-1}$Mpc of 1.6\% and therefore means that   
finding a galaxy number deficiency of 25\% over our sphere is a 15.6$\sigma$
result, although it is difficult to judge how much the $a\  posteriori$ selection
is affecting this result. 

However, if we assume our second case where r$_{cut}$=150$h^{-1}$Mpc then we find
an expected galaxy number fluctation of 8\%. This is a 3.75$\sigma$ result and
so a power law correlation function extending to 150\mpc\  is starting to be more consistent
with the observation of the local underdensity.

We take the $\Lambda$CDM real-space correlation function of Padilla \& Baugh (2003), which
we have modelled by two power laws of $\gamma$=-1.7 and $\gamma$=-3.44 with a break at r$_{break}$=30\mpc. 
This overestimates the $\Lambda$CDM $\xi(r)$ at all scales for r$\gsim100$\mpc. Even so, integrating
this model to r$_s$=150\mpc, we find that a 1$\sigma$ fluctuation is of order 4.8\% indicating that
our 25\% observed fluctuation is a 5$\sigma$ deviation. 

We conclude that there may be more power required at large scales than in the $\Lambda$CDM $\xi(r)$; 
although biasing the galaxy distribution could improve the agreement, in standard versions of the
$\Lambda$CDM model the galaxy distribution is predicted to be unbiased on large scales. 
The implied excess power is also in contradiction with the currently observed forms for the APM and 
2dFGRS correlation functions (Hawkins et al. 2002). Either these correlation functions have underestimated 
the amount of power at large scales or the galaxy distribution in the SGC is atypical of the
overall distribution of galaxies. We are therefore currently reanalysing the correlation functions of the
2dFGRS to check whether any such problem exists (Frith et al., in preparation).
 
\section{Conclusions}
In this paper we first presented the resulting galaxy counts from our CTIO
Curtis Schmidt data.  The large deficiency of galaxies seen in the SGC motivated
us to investigate the possible existence of a large ``hole'' in the SGC galaxy
distribution.  Using our CTIO data, covering 300 deg$^2$ in the SGC, we were able
to make the first ever detailed checks of the bright (B$<$17) galaxy photometry
in the DUKST, 2dFGRS and APM surveys which has crucial implications for the existence of
a local hole in the distribution of galaxies.  Our conclusions are:
\begin{itemize}

\item{Our B-band galaxy counts in the NGC agree extremely well with the model of
Metcalfe et al. (2001) but our SGC counts shows a significant galaxy deficiency, 
a mean 30.7\% in the magnitude range 14$<$B$<$18.5.}
\item{Good agreement is found for our NGC data with the SDSS and MGC number
counts in the magnitude interval 16.5$<$B$<$18.5 and likewise for our SGC galaxy 
counts with the data of Bertin \& Dennefeld and the APM survey in the range   
16.5$<$B$<$19. }
\item{We compared our CCD galaxy catalogue to that of the 2dFGRS in the NGC and
SGC.  In the NGC we find good agreement of the zero-point at b$_J$=16 but our    
galaxies are, on average, 0.13 mag. brighter at b$_J$=18 implying a scale error
of 0.065 mag/mag.  We find a mean zero-point difference of 0.1 mag. in the range
16$<$b$_J$$<$18.  In the SGC we find no zero-point or scale errors.}
\item{After applying this 0.1 mag. zero-point correction to the 2dF NGC
photometry we then find excellent agreement in both galactic caps between our
CTIO data and the 2dFGRS, who's zero-point corrected results also imply a 30\%
normalisation difference between the NGC and SGC galaxy counts at B=18.} 
\item{After comparing our CTIO photometry to both the DUKST and APMBGC data we 
found that our galaxies were, on average, brighter by 0.24 mag. and 0.31 mag.   
respectively.}
\item{Our R-band galaxy counts show a normalisation difference for the NGC and 
SGC data of $\sim$22.2\% in the range 13$<$R$<$17.}
\item{The 2dF and DUKST n(z) distributions show striking common structure with
regard to 2 ``holes'' in the galaxy distribution in the redshift ranges
0.03$<$z$<$0.05 and 0.06$<$z$<$0.09, with number discrepancy percentages of
35-50\% , 25-60\% respectively.  }
\item{The DUKST survey finds significantly larger galaxy number discrepancies   
for z$>$0.06 than the 2dF survey in the SGC.  This
is evidence that bright galaxies may be biased tracers of the underlying mass
distribution.}
\item{Using variable $\phi^{\star}$ models based on the structure seen in the
DUKST and 2dFGRS n(z)'s we claim that the galaxy count data is consistent with
there being a hole in the SGC galaxy distribution over the whole 4300 deg$^2$ APM
area, which extends out to z=0.1.}
\item{Taking together the deficiency in the APMBGC counts and our variable
$\phi^{\star}$ models we conclude that there is a galaxy number deficiency of $\approx$25\%
over a huge angular area of 100$^{\circ}$x60$^{\circ}$ in the SGC.  The evidence 
is that this hole extends out  to z=0.1 or 300$h^{-1}$Mpc. This result agrees well with
those of Frith et al. (2003) }
\item{We show, using a simple model, that significant power is required in the
two-point correlation function on very large scales if there is to be any
reasonable chance of the existence of such a large hole in the galaxy
distribution. The $\Lambda$CDM model shows less excess power than required to explain
the existence of such a feature in the galaxy distribution unless it is significantly 
biased on $\approx$150\mpc\  scales. This is not expected in current $\Lambda$CDM scenarios which
are supposed to have b=1 for r\gsim1\mpc. }

\end{itemize}

\end{document}